\documentclass[conference]{IEEEtran}
\IEEEoverridecommandlockouts
\usepackage{cite}
\usepackage{amsmath,amssymb,amsfonts,subcaption}
\usepackage{algorithmic}
\usepackage{graphicx}
\usepackage{textcomp}
\usepackage{xcolor}
\usepackage[colorlinks,
            linkcolor=red,
            anchorcolor=blue,
            citecolor=green]{hyperref}
\usepackage{amsmath,graphicx,pifont,multirow,booktabs,bbding,array,cleveref,amssymb,float,hyperref}
\def\BibTeX{{\rm B\kern-.05em{\sc i\kern-.025em b}\kern-.08em
    T\kern-.1667em\lower.7ex\hbox{E}\kern-.125emX}}
\begin{document}

\title{OCTAMamba: A State-Space Model Approach for Precision OCTA Vasculature Segmentation \\
\thanks{}
}

\author{
	\IEEEauthorblockN{
		Shun Zou\IEEEauthorrefmark{1,4}$^{\ast}$\thanks{$^{\ast}$Equal Contribution.}, 
		Zhuo Zhang\IEEEauthorrefmark{2}$^{\ast}$, 
		Guangwei Gao\IEEEauthorrefmark{3,4}$^{\dagger}$\thanks{$^{\dagger}$Corresponding author.}\thanks{This work was supported in part by the Open Fund Project of Provincial Key Laboratory for Computer Information Processing Technology (Soochow University) under Grant KJS2274.}} 
	\IEEEauthorblockA{\IEEEauthorrefmark{1}College of Artificial Intelligence, Nanjing Agricultural University, Nanjing, China}
	\IEEEauthorblockA{\IEEEauthorrefmark{2}College of Computer Science and Technology, National University of Defense Technology, Changsha, China}
	\IEEEauthorblockA{\IEEEauthorrefmark{3}Institute of Advanced Technology, Nanjing University of Posts and Telecommunications, Nanjing, China}
 \IEEEauthorblockA{\IEEEauthorrefmark{4}Provincial Key Laboratory for Computer Information Processing Technology, Soochow University, Suzhou, China}

	\IEEEauthorblockA{
 zs@stu.njau.edu.cn, zhangzhuo@nudt.edu.cn, csggao@gmail.com}
}

\DeclareRobustCommand*{\IEEEauthorrefmark}[1]{%
    \raisebox{0pt}[0pt][0pt]{\textsuperscript{\footnotesize\ensuremath{#1}}}}


\maketitle

\begin{abstract}
Optical Coherence Tomography Angiography (OCTA) is a crucial imaging technique for visualizing retinal vasculature and diagnosing eye diseases such as diabetic retinopathy and glaucoma. However, precise segmentation of OCTA vasculature remains challenging due to the multi-scale vessel structures and noise from poor image quality and eye lesions. In this study, we proposed OCTAMamba, a novel U-shaped network based on the Mamba architecture, designed to segment vasculature in OCTA accurately. OCTAMamba integrates a Quad Stream Efficient Mining Embedding Module for local feature extraction, a Multi-Scale Dilated Asymmetric Convolution Module to capture multi-scale vasculature, and a Focused Feature Recalibration Module to filter noise and highlight target areas. Our method achieves efficient global modeling and local feature extraction while maintaining linear complexity, making it suitable for low-computation medical applications. Extensive experiments on the OCTA 3M, OCTA 6M, and ROSSA datasets demonstrated that OCTAMamba outperforms state-of-the-art methods, providing a new reference for efficient OCTA segmentation. Code is available at \href{https://github.com/zs1314/OCTAMamba}{https://github.com/zs1314/OCTAMamba}
\end{abstract}


\section{Introduction}
\label{sec:intro}
Optical coherence tomography angiography (OCTA) is a non-invasive imaging technique that provides detailed visualization of the retinal vasculature \cite{kashani2017optical} \cite{spaide2018optical}. Segmenting OCTA vasculature aids in diagnosing various eye diseases, such as diabetic retinopathy (DR) \cite{PAN2018146} and glaucoma \cite{jia2014optical}. Therefore, developing an automated OCTA vasculature segmentation model is crucial for improving the diagnostic efficiency of eye diseases.

In recent years, convolutional neural networks (CNNs) have been widely used for OCTA image segmentation \cite{li2020image,li2022image,hu2022joint}. For example, UNet's symmetric encoder-decoder structure and skip connections extract features at different levels, enabling efficient feature transformation and laying the foundation for medical segmentation. This has led to many derivative works based on the U-shaped structure. Ma et al \cite{ma2020rose} proposed attention residual UNet to refine segmentation results further. Ziping et al \cite{ma2022retinal} introduced a contrastive learning module to improve UNet's performance. However, these CNN-based models have limitations in modeling long-range dependencies due to their limited receptive fields. In contrast, Transformer-based models effectively capture global information through self-attention mechanisms and perform remote spatial modeling \cite{dosovitskiy2020vit}. However, their quadratic complexity related to image size results in substantial computational costs, especially in pixel-dense prediction tasks such as medical image segmentation \cite{zhang2023novel}\cite{perslev2019one}. This does not meet the requirements for lightweight models needed in mobile medical tasks with low parameters and low computational complexity.

Recently, structured state-space sequence models (SSMs) \cite{kalman1960new,orvieto2023resurrecting,zou2024skinmamba,ge2024mambatsr,gu2021combining}, such as Mamba \cite{gu2023mamba}, have emerged as powerful methods for long-sequence modeling, achieving effective global modeling with linear complexity. For example, VM-Unet \cite{ruan2024vmunetvisionmambaunet} introduces the Visual State Space (VSS) module as a foundational component to capture extensive contextual information through an asymmetric encoder-decoder structure. U-Mamba \cite{U-Mamba} incorporates an SSM-Conv hybrid module, leveraging CNN's local feature extraction capabilities and Mamba's long-range modeling strengths. Mamba-UNet \cite{wang2024mamba} is a novel architecture that integrates the strengths of U-Net with the Mamba architecture, leveraging its ability to model long-range dependencies for improved medical image segmentation. The model demonstrates superior performance on cardiac MRI and abdominal CT datasets compared to other U-Net variants. However, these methods still have limitations as they do not consider some critical characteristics of OCTA images. For instance, the retinal vasculature in OCTA images is multi-scaled, and fine, terminal branches are challenging to segment accurately. Additionally, poor image quality, incorrect layer projection, and eye lesions introduce noise, adversely affecting precise retinal vasculature segmentation.

\begin{figure*}[htp]
\centering
\includegraphics[width=\linewidth ]{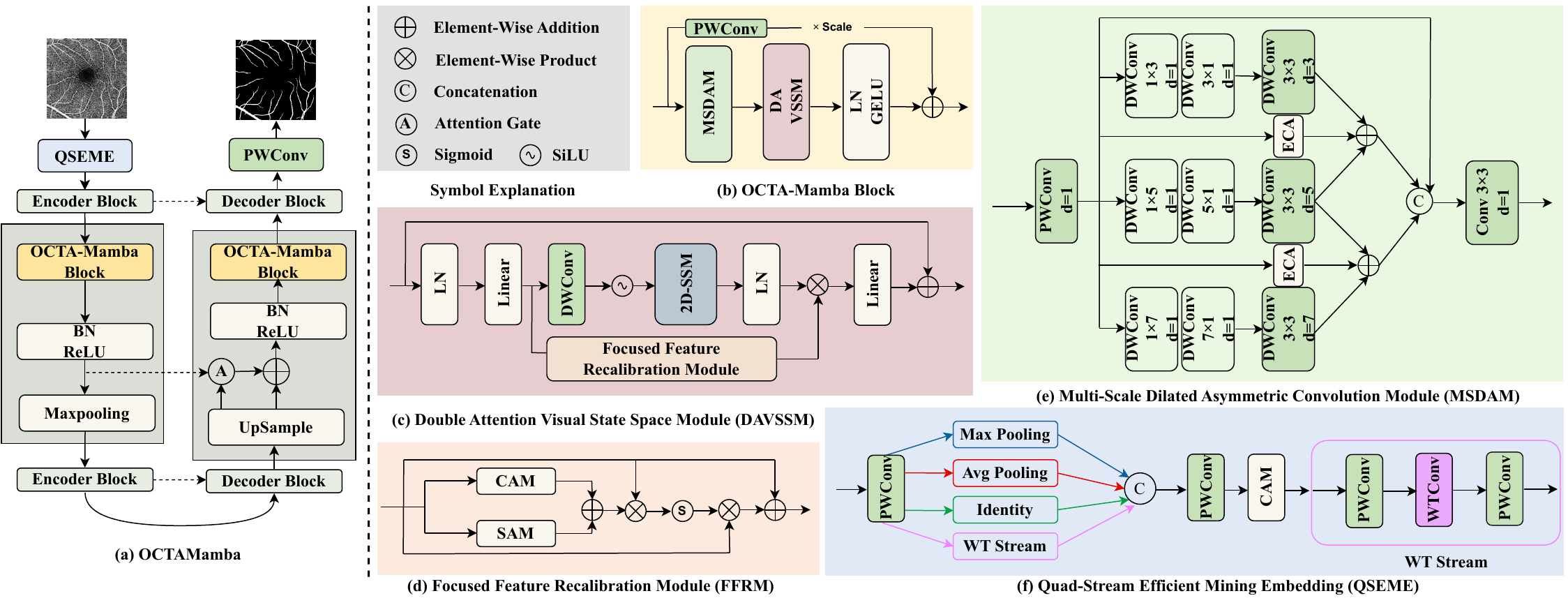}
\caption{Overall and detailed architecture of the OCTAMamba.}
\label{fig1}
\vspace{-0.5cm}
\end{figure*}

To address these issues, we meticulously designed a U-shaped network for OCTA vasculature segmentation based on the Mamba architecture: OCTAMamba. This network effectively captures local information and models long-range dependencies while extracting multi-scale information from OCTA images to enhance feature representation and filter out noise to highlight target areas. Additionally, it operates efficiently in low-computation medical scenarios due to its extremely low parameter count. The main contributions of this paper are as follows:
\begin{itemize}
\item We designed Quad Stream Efficient Mining Embedding (QSEME) as a pre-component of the Mamba module to supplement local features. We also proposed the Multi-Scale Dilated Asymmetric Convolution Module (MSDAM) to capture vasculature at various scales, including fine, terminal branches. Additionally, we introduced the Focused Feature Recalibration Module (FFRM), which is highly integrated with the VSS module to recalibrate and filter features, reducing noise interference for precise vasculature segmentation, capturing complex spatial relationships, and focusing on significant regions.

\item To the best of our knowledge, we are the first to successfully introduce Mamba into the OCTA image segmentation task, providing a new reference for future efficient OCTA vasculature segmentation exploration.

\item We conducted extensive experiments on the OCTA\_3M, OCTA\_6M, and ROSSA datasets. The results demonstrated that our proposed method outperformed other state-of-the-art methods.
\end{itemize}

\section{Method}
\label{sec:method}
\Cref{fig1} \textcolor{red}{(a)} illustrates the overall architecture of the proposed OCTAMamba, which follows an encoder-decoder structure capable of effectively capturing local features and global information. Specifically, OCTAMamba initially inputs the OCTA image into Quad Stream Efficient Mining Embedding (QSEME) for multi-stream perception feature extraction, followed by three consecutive encoder blocks to further extract features. Each encoder block primarily consists of an OCTA-Mamba Block, Batch Normalization (BN), ReLU activation function, and maxpooling for downsampling. With each encoder block passed, the height and width of the input features are halved, while the number of channels doubles. Similarly, the decoder is composed of three consecutive decoder blocks, mainly featuring Upsample for upsampling, BN, ReLU activation function, and OCTA-Mamba Block. Each decoder block restores the image size to twice its previous state while halving the number of channels. After the decoder, we use point-wise convolution (PWConv) to restore the channel count to match the segmentation target. Regarding skip connections, we introduce an Attention Gate \cite{oktay2018attentionunetlearninglook}, which does not require high computational costs but significantly enhances the quality of the skip connections.
QSEME and OCTA-Mamba Block are the core components of our proposed network, detailed further in \Cref{ssec:QSEME} and \Cref{ssec:octa-mamba}.

\begin{table*}[htp]
    \centering
      \caption{Performance comparison of different methods on three public datasets. The best results are highlighted in \textbf{bold fonts}. “ ↑ ”and “ ↓ ” indicate that larger or smaller is better.}\begin{tabular}{lc>{\centering\arraybackslash}p{1.1cm}ccccccccc}
        \toprule
        \multirow{2}{*}{\textbf{Method}} & \multirow{2}{*}{\textbf{Year}}  &\multirow{2}{*}{\textbf{Params} $\downarrow$}& \multicolumn{3}{c}{\textbf{OCTA\_3M} \cite{li2022octa500retinaldatasetoptical}} & \multicolumn{3}{c}{\textbf{OCTA\_6M} \cite{li2022octa500retinaldatasetoptical}} & \multicolumn{3}{c}{\textbf{ROSSA} \cite{ning2023accurate}} \\
        \cmidrule(lr){4-12}
         & & & \textbf{Dice} $\uparrow$ & \textbf{IoU} $\uparrow$& \textbf{Sen} $\uparrow$ & \textbf{Dice} $\uparrow$ & \textbf{IoU} $\uparrow$& \textbf{Sen} $\uparrow$ & \textbf{Dice} $\uparrow$ & \textbf{IoU} $\uparrow$& \textbf{Sen} $\uparrow$ \\
        \midrule
        U-Net \cite{ronneberger2015u}& MICCAI-2015  & 15.04& 79.36& 65.86& 79.81& 77.32& 63.11& 78.88& 83.82& 72.25& 84.76\\
        R2Unet \cite{Alom2018RecurrentRC} & Arxiv-2018  & 39.09& 66.32& 46.31 & 69.33& 51.78& 35.16& 44.82& 78.27& 64.42&79.21\\
        UNet++ \cite{zhou2019unetplusplus} & TMI-2019 &26.9& 82.83& 70.86& 82.31& 79.60& 66.21& 80.06& 88.88& 80.14& 87.57\\
        Swin-UNet \cite{swinunet} & ECCVW-2022 & 41.38& 73.66 & 58.40& 72.01& 72.39 & 56.84 &72.10& 79.19 & 65.68 &76.86\\
        H2Former \cite{10093768} & TMI-2023 & 33.67& 80.13 & 66.90 & 82.32& 74.19 & 59.04 & 78.07& 83.74 & 72.13& 83.37\\
        MISSFormer \cite{9994763} & TMI-2023 & 42,46& 80.63 & 67.62& 80.72& 78.03 & 64.07& 78.97& 84.27 & 72.93& 85.34\\
        U-Mamba \cite{U-Mamba} & Arxiv-2024 & 18.39& 75.76 & 61.07 &77.56& 70.24 & 54.22& 73.41& 77.46 & 63.33& 80.43\\
        VM-UNet \cite{ruan2024vmunetvisionmambaunet} & Arxiv-2024 & 44.27& 71.98 & 56.34 &69.67& 71.55 & 55.81 &70.54& 81.10 & 68.33& 81.64\\
        AC-Mamba \cite{Nguyen2024ACMAMBASEGAA}& Arxiv-2024 & 7.99& 80.44 & 67.36& 79.13& 78.42 & 64.61 &77.19& 88.85 & 80.10& 87.57\\
        H-vmunet \cite{wu2024h}& Arxiv-2024  & 8.97& 67.15& 50.66&69.10& 64.18& 47.36&66.61& 70.33& 54.38&71.10\\
        OCTAMamba & - & \textbf{3.57}& \textbf{84.50}& \textbf{73.23}&\textbf{84.00}& \textbf{82.31}& \textbf{70.03}&\textbf{82.75}& \textbf{90.04}& \textbf{82.03}&\textbf{88.86}\\
        \bottomrule
    \end{tabular}
  
    \label{tab:octa_comparison}
\vspace{-0.5cm}
\end{table*}
\subsection{Quad Stream Efficient Mining Embedding }
\label{ssec:QSEME}
As shown in \Cref{fig1} \textcolor{red}{(f)}, the Quad Stream Efficient Mining Embedding (QSEME) consists of max pooling perceived flow (mppf), average pooling perceived flow (appf), residual perceived flow (rpf), and wavelet transform perceived flow (wtpf). The wtpf is the core component of QSEME, designed to efficiently expand the receptive field by focusing on different frequency bands of the OCTA image through WTConv \cite{finder2024wavelet}. It is composed of two PWConv and WTConv layers. First, the input features pass through PWConv to increase the number of channels, and then the feature map is fed into the four perceived flows. Finally, we integrate the feature maps output by the four perceived flows through concatenation, adjust the channel count with PWConv, and enhance the interaction of channel information and local information using the Channel Attention Module (CAM). Mathematically, the above process could be elaborated as: 
 \begin{gather}
    f={PW}(x) 
  \\
      f_{wtpf}=PW(WTConv(PW(f)))
 \\
      f_{mppf}=\wp_{Max}(f)
\\
  f_{appf}=\wp_{Avg}(f)
  \\
    f_{rpf}=f
\\
F_{QSEME}=CAM(PW([f_{wtpf},f_{mppf},f_{appf},f_{rpf}]))
\end{gather}
where \( \text{PW}(\cdot) \) denotes the point-wise convolution, \( \wp_{\text{Max}}(\cdot) \) represents the max pooling operation, \( \wp_{\text{Avg}}(\cdot) \) represents the average pooling operation, \([ \cdot, \cdot ]\) indicates the concatenation operation, \( \text{CAM}(\cdot) \) denotes the Channel Attention Module and \( F_{QSEME} \) represents the final output of this module.

\subsection{OCTA-Mamba Block }
\label{ssec:octa-mamba}
As shown in \Cref{fig1} \textcolor{red}{(b)}, the core of the OCTA-Mamba Block consists of the Multi-Scale Dilated Asymmetric Convolution Module (MSDAM) and the Double Attention Visual State Space Module (DAVSSM). Additionally, the scaled residuals obtained by multiplying the input features with the learnable scaling parameters help maintain consistent information flow before and after the OCTA-Mamba Block. Layer Normalization (LN) and GELU activation functions enhance the model's nonlinearity and stability.

\textbf{Multi-Scale Dilated Asymmetric Convolution Module (MSDAM).} The MSDAM is a unique six-branch structure. Three branches capture multi-scale information through different sizes of convolution, obtaining scale features. Two middle branches preserve input information via ECA \cite{wang2020ecanetefficientchannelattention}, resulting in retention features. The final residual branch aids in promoting gradient propagation, yielding residual features. We employ asymmetric convolution and depth-wise separable convolution (DWConv) to effectively reduce network parameters and use dilated convolution to expand the receptive field. At the end of the module, features are divided into two groups, each containing two scale features and one retention feature, which are then fused. Finally, each group produces two multi-scale feedforward features, which are concatenated with residual features. The output feature map is fed into the last convolution layer to restore the final channel dependency. The entire operation of the module is defined as follows:
\begin{gather}
    x=f_{d=1}^{1\times1}(F)
    \\
    F_{scale}^{i}=DW_{d=i}^{3\times 3}(DW_{d=1}^{i\times 1}((DW_{d=1}^{1\times i}(x))), i=3,5,7
    \\
    F_{retention}^{1}=F_{retention}^{2}=ECA(x)
    \\
    F_{residual}=x
    \\
     Group_{1}=F_{scale}^{3} \otimes  F_{retention}^{1}\otimes F_{scale}^{5} 
     \\
          Group_{2}=F_{scale}^{5} \otimes  F_{retention}^{2}\otimes F_{scale}^{7} 
        \\
    F_{MSDAM}=f_{d=1}^{3 \times 3}([Group_{1},Group_{2},  F_{residual}])
\end{gather}
where  \( f^{x \times y} (\cdot)\) denotes the \( X \times Y \) standard convolution operation, \( DW^{x \times y}  (\cdot)\) denotes the \( X \times Y \) depth-wise separable convolution operation, the parameter \( d \) represents the dilation rate of the dilated convolution, \( ECA (\cdot) \) denotes the Efficient Channel Attention mechanism, \( \otimes \) denotes element-wise product and \( F_{MSDAM} \) represents the final output of this module.

\textbf{Double Attention Visual State Space Module (DAVSSM).} \Cref{fig1} \textcolor{red}{(c)} describes the DAVSSM. The input features are normalized and then split into two branches after passing through the Linear layer. 
This process is represented as:
\begin{equation}
        F_{1}, F_{2}={Linear}({LN(F_{MSDAM})}) 
\end{equation}
where \( F_{MSDAM} \) represents the output of MSDAM, \( \text{LN}( \cdot ) \) denotes the layer normalization operation, and \( \text{Linear}( \cdot ) \) denotes the processing using a linear layer. \( F_1 \) and \( F_2 \) are the inputs for the first and second branches, respectively. In the first branch, \( F_1 \) is processed through DWConv, SiLU, and the 2D selective scanning (SS2D) module for further feature extraction, followed by normalization to output \( F_{{extract}} \). In the second branch, \( F_2 \) undergoes the Focused Feature Recalibration Module for feature selection, outputting \( F_{{select}} \). Finally, the outputs of both branches are element-wise multiplied, passed through a Linear layer, and connected with the residual to obtain the module's output \( F_{{DAVSSM}} \). The entire operation of the module is defined as follows:
\begin{gather}
    F_{extract}={LN}({SS2D}({DW(F_{1})})) 
    \\
    F_{select}={FFRM}(F_{2})
    \\
    F_{DAVSSM}={Linear}(F_{extract}\otimes F_{select})\oplus F_{MSDAM} 
\end{gather}
where \( \text{SS2D}( \cdot ) \) is 2D selective scanning, and \( \text{FFRM}( \cdot ) \) denotes processing through the Focused Feature Recalibration Module, which will be detailed in the next subsection. \( F_{{extract}} \) and \( F_{{select}} \) are the output features of the first and second branches, respectively. \( \oplus \) represents element-wise addition and \( F_{{DAVSSM}} \) is the output of the DAVSSM.

\begin{figure*}[h]
    \centering
     
    \begin{subfigure}[b]{0.12\linewidth}
        \captionsetup{labelformat=empty}
        \captionsetup{skip=2pt} 
        \includegraphics[scale=0.35]{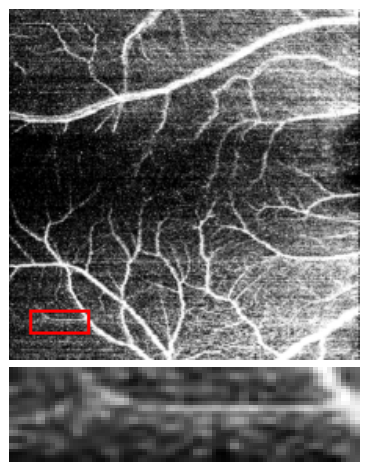}
        \caption{\scriptsize Original}
    \end{subfigure}
    \begin{subfigure}[b]{0.12\linewidth}
        \captionsetup{labelformat=empty}
        \captionsetup{skip=2pt} 
        \includegraphics[scale=0.35]{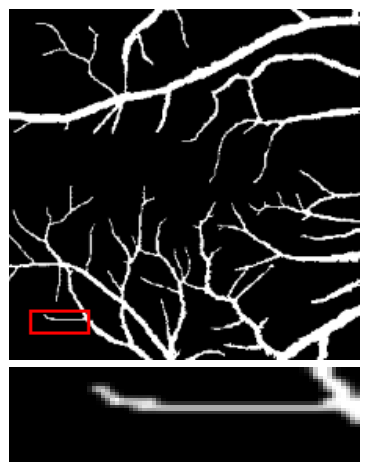}
        \caption{\scriptsize GT}
    \end{subfigure}
    \begin{subfigure}[b]{0.12\linewidth}
        \captionsetup{labelformat=empty}
        \captionsetup{skip=2pt} 
        \includegraphics[scale=0.35]{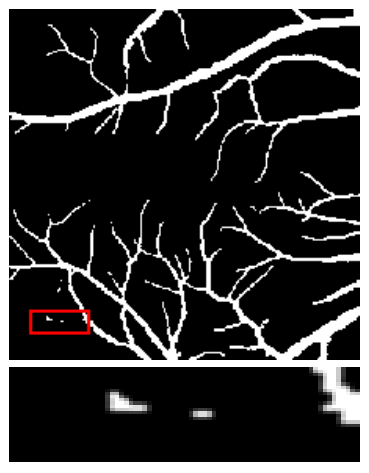}
        \caption{\scriptsize U-Net \cite{ronneberger2015u}}
    \end{subfigure}
    \begin{subfigure}[b]{0.12\linewidth}
        \captionsetup{labelformat=empty}
        \captionsetup{skip=2pt} 
        \includegraphics[scale=0.35]{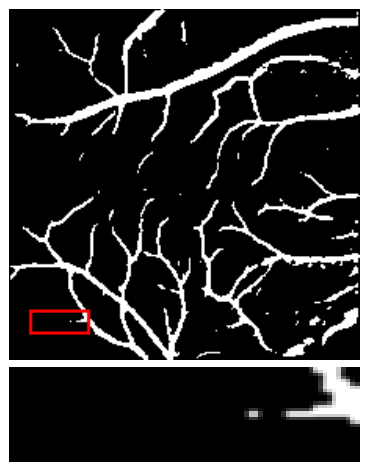}
        \caption{\scriptsize R2Unet \cite{Alom2018RecurrentRC}}
    \end{subfigure}
    \begin{subfigure}[b]{0.12\linewidth}
        \captionsetup{labelformat=empty}
        \captionsetup{skip=2pt} 
        \includegraphics[scale=0.35]{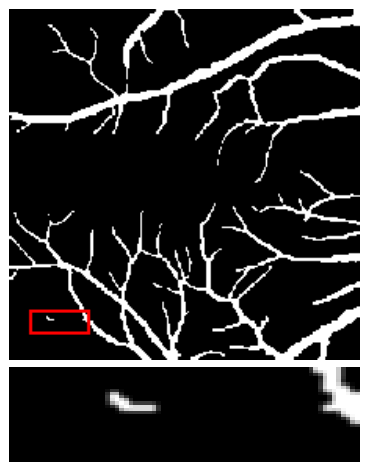}
        \caption{\scriptsize UNet++ \cite{zhou2019unetplusplus}}
    \end{subfigure}
    \begin{subfigure}[b]{0.12\linewidth}
        \captionsetup{labelformat=empty}
        \captionsetup{skip=2pt} 
        \includegraphics[scale=0.35]{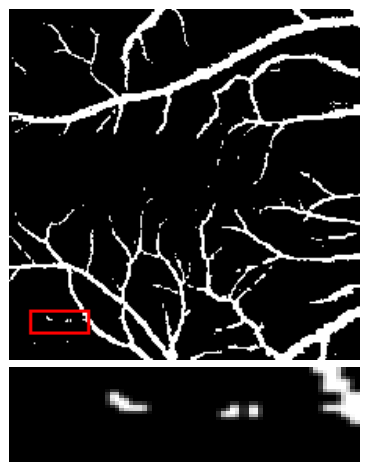}
        \caption{\scriptsize Swin-UNet \cite{swinunet}}
    \end{subfigure}

    \begin{subfigure}[b]{0.12\linewidth}
        \captionsetup{labelformat=empty}
        \captionsetup{skip=2pt} 
        \includegraphics[scale=0.35]{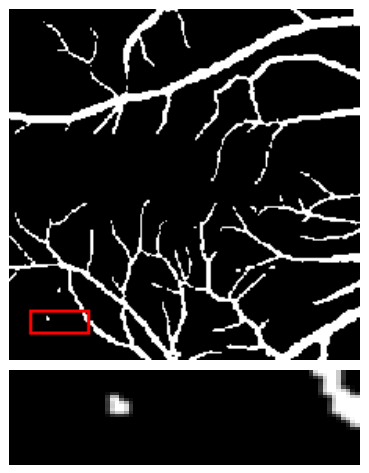}
        \caption{\scriptsize H2Former \cite{10093768}}
    \end{subfigure}
    \begin{subfigure}[b]{0.12\linewidth}
        \captionsetup{labelformat=empty}
        \captionsetup{skip=2pt} 
        \includegraphics[scale=0.35]{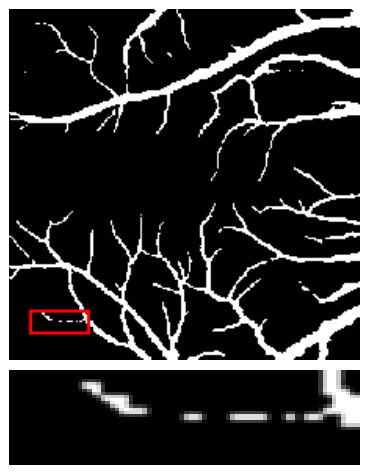}
        \caption{\scriptsize MISSFormer \cite{9994763}}
    \end{subfigure}
    \begin{subfigure}[b]{0.12\linewidth}
        \captionsetup{labelformat=empty}
        \captionsetup{skip=2pt} 
        \includegraphics[scale=0.35]{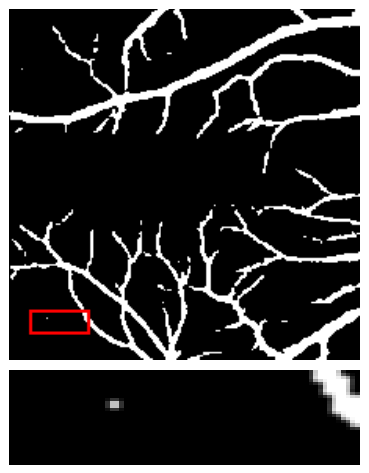}
        \caption{\scriptsize U-Mamba \cite{U-Mamba}}
    \end{subfigure}
    \begin{subfigure}[b]{0.12\linewidth}
        \captionsetup{labelformat=empty}
        \captionsetup{skip=2pt} 
        \includegraphics[scale=0.35]{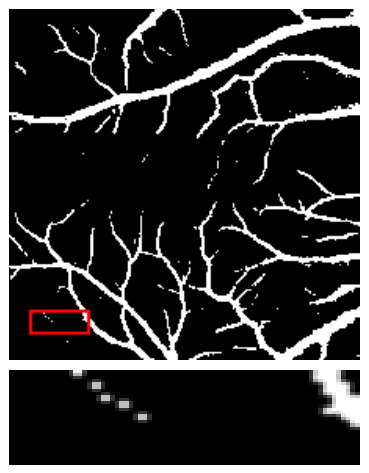}
        \caption{\scriptsize VM-UNet \cite{ruan2024vmunetvisionmambaunet}}
    \end{subfigure}
    \begin{subfigure}[b]{0.12\linewidth}
        \captionsetup{labelformat=empty}
        \captionsetup{skip=2pt} 
        \includegraphics[scale=0.35]{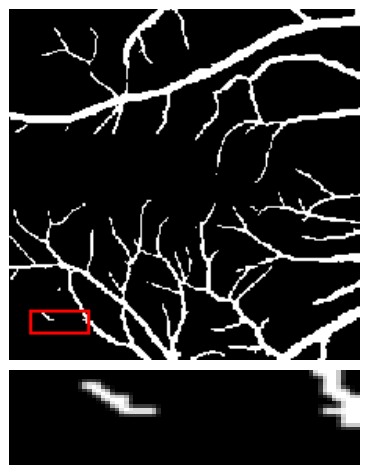}
        \caption{\scriptsize AC-Mamba \cite{Nguyen2024ACMAMBASEGAA}}
    \end{subfigure}
    \begin{subfigure}[b]{0.12\linewidth}
        \captionsetup{labelformat=empty}
        \captionsetup{skip=2pt} 
        \includegraphics[scale=0.35]{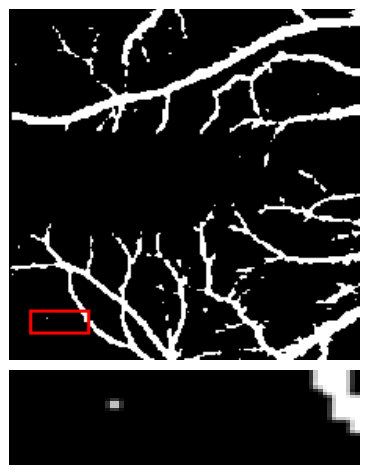}
        \caption{\scriptsize H-vmunet \cite{wu2024h}}
    \end{subfigure}
    \begin{subfigure}[b]{0.12\linewidth}
        \captionsetup{labelformat=empty}
        \captionsetup{skip=2pt} 
        \includegraphics[scale=0.35]{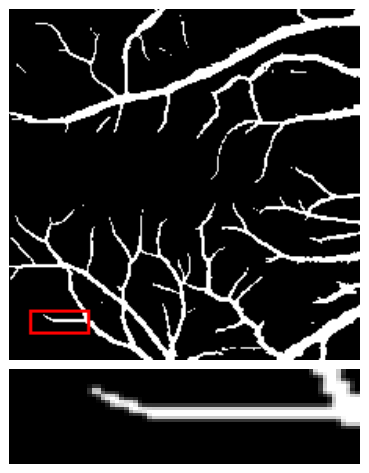}
        \caption{\scriptsize Ours}
    \end{subfigure}

    \caption{Qualitative visualization of different methods. Best viewed by zooming in the figures on high-resolution displays.}
    \label{fig:methods_comparison}
    \vspace{-0.5cm}
    \label{fig2}
\end{figure*}



\textbf{Focused Feature Recalibration Module (FFRM).} As shown in \Cref{fig1} \textcolor{red}{(d)}, we proposed a Focused Feature Recalibration Module combining channel and spatial attention to further select features from the Mamba output, highlighting the tiny target vessels in OCTA images. The input feature map is processed in parallel through the Channel Attention Module (CAM) and the Spatial Attention Module (SAM). The outputs of these two branches are fused to calibrate the input feature map for the first time, followed by the Sigmoid function for the second feature selection. Finally, a residual structure is introduced to accelerate model convergence.

\section{EXPERIMENT}
\label{sec:experiment}

\subsection{Dataset and Implementation Details}
\label{ssec:data and details}

In our experiments, we used the publicly available OCTA-500 \cite{li2022octa500retinaldatasetoptical} and ROSSA \cite{ning2023accurate} datasets. OCTA-500 is divided into two subsets: OCTA\_6M with 300 samples and OCTA\_3M with 200 samples. ROSSA contains 618 samples. The training, validation, and test sets for all datasets were split according to the literature \cite{ning2023accurate}. Additionally, the data underwent normalization and standardization. We used the Dice similarity coefficient (Dice), Intersection over Union (IoU), Sensitivity (Sen), and model params as the primary evaluation metrics.

We implemented our experiment using PyTorch 2.0.0 and trained it on a single NVIDIA V100 Tensor Core GPU (32GB) for 400 epochs with a batch size of 2. The input images were uniformly resized to 224 × 224. We employed the AdamW optimizer with an initial learning rate of 0.0001 and a weight decay of 0.001, and used DiceLoss to optimize the model parameters. Additionally, we applied an early stopping strategy. To ensure the reproducibility of the experimental results, we fixed the random seed to 0.
\begin{table}[t]
    \centering
     \caption{Comparison of different versions with various modules and their performance on OCTA\_3M \cite{li2022octa500retinaldatasetoptical}.}\begin{tabular}{
     >{\centering\arraybackslash}p{0.3cm}
     >{\centering\arraybackslash}p{1.1cm}
     >{\centering\arraybackslash}p{1.1cm}
     >{\centering\arraybackslash}p{1.1cm}
     ccc}
        \toprule
        \multirow{2}{*}{\textbf{Ver.}} & \multirow{2}{*}{\textbf{QSEME}} & \multirow{2}{*}{\textbf{MSDAM}}& \multirow{2}{*}{\textbf{FFRM}}& \multicolumn{3}{c}{\textbf{OCTA\_3M} \cite{li2022octa500retinaldatasetoptical}} \\
        \cmidrule(lr){5-7}
         & & & & \textbf{Dice} & \textbf{IoU}& \textbf{Sen} \\
        \midrule
        No.1 & - & - & - &  82.01& 69.12& 82.96\\
        No.2 & \checkmark & - & - & 82.98& 70.25& 83.78\\
        No.3 & - & \checkmark & - & 83.06& 71.47& 83.92\\
        No.4 & - & - & \checkmark & 82.93& 72.37&83.12\\
        No.5 & \checkmark & \checkmark & - & 84.01& 71.99&83.96\\
        No.6 & \checkmark & - & \checkmark & 83.92& 72.36&83.93\\
        No.7 & - & \checkmark & \checkmark & 84.02& 72.42&83.95\\
        No.8 & \checkmark & \checkmark & \checkmark & \textbf{84.50}& \textbf{73.23}&\textbf{84.00}\\
        \bottomrule
    \end{tabular}
   
    \label{tab:ablation}
\vspace{-0.2cm}
\end{table}

\subsection{Comparison with the State-of-the-art Methods    }
\label{ssec:compar}
To demonstrate the effectiveness of our method, we compared OCTAMamba with three other types of methods: CNN-based methods (UNet \cite{ronneberger2015u}, R2Unet \cite{Alom2018RecurrentRC} and UNet++ \cite{zhou2019unetplusplus}), Transformer-based methods (Swin-UNet \cite{swinunet}, H2Former \cite{10093768} and MISSFormer \cite{9994763}), and the latest Mamba-based segmentation networks (U-Mamba \cite{U-Mamba}, VM-UNet \cite{ruan2024vmunetvisionmambaunet}, AC-Mamba \cite{Nguyen2024ACMAMBASEGAA} and H-vmunet \cite{wu2024h}). Numerical results and visual effects are shown in \Cref{tab:octa_comparison} and \Cref{fig2}. As seen in \Cref{tab:octa_comparison}, our OCTAMamba outperformed other state-of-the-art segmentation methods in Dice, Iou, Sen, and Params metrics. As shown in \Cref{fig2}, OCTAMamba's results were closer to ground truth and achieved good segmentation results even in fine terminal branches.

\subsection{Ablation Experiments  }
\label{ssec:Ablation Experiments}
As shown in \Cref{tab:ablation}, to explore the impact of each component on model performance, we conducted ablation experiments on ROSSA\_3M. From \Cref{tab:ablation}, it is evident that QSEME, MSDAM, and FFRM all enhanced the model's segmentation of the target area to varying degrees. The performance of OCATMamba was optimal when all three modules were used simultaneously.

\section{ Conclusion }
\label{sec:Conclusion}
In this paper, we introduced OCTAMamba, an advanced network architecture designed for the efficient and precise segmentation of OCTA vasculature. By leveraging the strengths of Mamba architecture, we developed innovative modules such as the Quad Stream Efficient Mining Embedding, Multi-Scale Dilated Asymmetric Convolution Module, and Focused Feature Recalibration Module. These modules collectively enhance multi-scale feature representation and effectively filter noise, addressing the challenges posed by small and noisy vascular structures in OCTA images.
Extensive experimental evaluations on the three datasets demonstrated that OCTAMamba achieves superior segmentation performance compared to existing state-of-the-art methods. 


\vfill\pagebreak

\bibliography{refs}

\end{document}